\documentclass[twocolumn,showpacs,preprintnumbers,amsmath,amssymb,superscriptaddress,nofootinbib]{revtex4}
\usepackage{epsfig}
\usepackage{psfrag}
\usepackage{amsfonts}
\usepackage{graphicx}
\usepackage{dcolumn}
\usepackage{bm}
\usepackage{appendix}
\newbox\pippobox

\begin{document}

\title{Nonperturbative Renormalization Group for the Landau-de Gennes Model}

\author{Bin Qin}
\email{qinbin@mails.ccnu.edu.cn}
\affiliation{Institute of Particle Physics (IOPP) and Key Laboratory of Quark and Lepton Physics (MOE),  Central China Normal University, Wuhan 430079, China}

\author{Defu Hou}
\email{houdf@mail.ccnu.edu.cn}
\affiliation{Institute of Particle Physics (IOPP) and Key Laboratory of Quark and Lepton Physics (MOE),  Central China Normal University, Wuhan 430079, China}

\author{Mei Huang}
\email{huangm@ihep.acac.edu.cn}
\affiliation{Institute of High Energy Physics, Chinese Academy of Sciences, Beijing 100049,  China}

\author{Danning Li}
\email{lidanning@jnu.edu.cn}
\affiliation{Department of Physics and Siyuan Labotorary, Jinan University, Guangzhou 510632, China}

\author{Hui Zhang}
\email{Mr.zhanghui@mails.ccnu.edu.cn}
\affiliation{Institute of Particle Physics (IOPP) and Key Laboratory of Quark and Lepton Physics (MOE),  Central China Normal University, Wuhan 430079, China}
\affiliation{Physics Department and Center for Exploration of Energy and Matter, Indiana University, 2401 N Milo B. Sampson Lane, Bloomington, IN 47408, USA}

\date{\today}

\begin{abstract}
We studied the nematic isotropic phase transition by applying the functional renormalization group to the Landau-de Gennes model. We derived the flow equations for the effective potential as well as  the cubic and quartic ``couplings'' and the anomalous dimension. We then solved the coupled flow equations on a grid using Newton Raphson method. A first order phase transition is observed. We also investigated the nematic isotropic puzzle (the NI puzzle) in this paper. We obtained the NI transition temperature difference ${T_c-T^*}=5.85K$ with sizable improvement over previous results.
\end{abstract}

\keywords{Nonperturbative Renormalization Group, Landau-De Gennes Model}

\maketitle

\section{Introduction}

The nematic isotropic (NI) phase transition has been an important topic of research over the past few decades~\cite{Grams:1986, Gennes1995, Singh:2000}. In uniaxial nematic liquid crystals the centers of gravity of the molecules have no long range order while all their axes point in roughly the same direction, described by the director, around which there exists complete rotational symmetry. When raising temperature, its order parameter abruptly drops to zero and becomes an isotropic phase. Thus the NI phase transition is of first order in nature. It can be described phenomenologically by the  Landau mean field scalar model with a cubic term~\cite{Gennes1995}. But it is relatively weak because only orientational order is lost and the latent heat is small~\cite{Singh:2000}. This also leads to large pretransition anomalies such as specific heat,  which indicates that it is  close to being second order. In the isotropic phase, although the order parameter vanishes on average, the molecules are still parallel to each other over a characteristic distance (the correlation length) which describes the average size of the range of correlations between the fluctuations.  In order to take these effects into account de Gennes proposed a tensor order parameter model, denoted as Landau-de Gennes Model~\cite{Gennes1995, Mukherjee1994}. This model also gains insight into the longstanding puzzle about the low value of $\frac{T_c-T^*}{T^*}\approx0.1 \%$, where $T_c$ is the nematic-isotropic phase  transition temperature and $T^*$ is interpreted as the temperature at which the light-scattering  intensity diverges in the supercooled isotropic phase. In experiments, it is shown that $T_c-T^*=1K$ ($\frac{T_c-T^*}{T^*}\approx0.3\%$) in the case of  nematic liquid crystal 8CB , which is much smaller than the usual theoretical model predictions. For instance the mean field result gives $T_c-T^*=24K$. By including fluctuations and using Wilson's renormalization group analysis, Mukherjee~\cite{Mukherjee1994, Mukherjee1995, Mukherjee1998} improved the result to be $T_c-T^*=7.47K$.

The nonperturbative renormalization group (NPRG, also called the functional renormalization group) \cite{Wetterich:1992yh} has been proven to be an extremely versatile and efficient tool to deal with fluctuations in recent years~\cite{Fukushima:2010ji, Jiang:2012wm, Rischke2013, Wjf2016, Delamotte:2016acs,  Eichhorn:2013zza, Fejos:2014qga,Delamotte:2003dw,Kopietz:2010zz, Berges:2000ew, JanP2007, Delamotte:2007pf,Metzner:2011cw},
see \cite{Kopietz:2010zz, Berges:2000ew, JanP2007, Delamotte:2007pf,Metzner:2011cw} for an excellent introduction. With this method one can systematically extract quantitative  reliable results about long distance physics  from short distance ansatz. As opposed to the usual functional integral approach in field theory, its conceptual framework is relatively simple and unified, and essentially takes the form of functional differential equations, which are more convenient for numerical computations. So the goal of this manuscript is to solve the Landau-de Gennes Model using the methods of nonperturbative renormalization group and improve the result of  NI puzzle in this framework.

The article is organized as follows. In Section \ref{sec-model}, we defined the model and notations. Then we give a quick overview of the FRG formalism and  apply it to the model. The concrete flow equations are derived, including the anomalous dimension. In Section \ref{sec-results}, we show our numerical results about effective potential. We also give our analysis of the nematic-isotropic puzzle. In Section \ref{sec-summary} we give our concluding remarks and outlook for future work.

\section{Application of Nonperturbative Renormalization Group to the Landau-de Gennes Model }
\label{sec-model}

The Lagrangian  density of the Landau-de Gennes model can be written as~\cite{Mukherjee1995,Mukherjee1994}
\begin{eqnarray}
L=\frac{1}{2} A\, \text{Tr}\left[Q^2\right]+\frac{1}{3} B\, \text{Tr}\left[Q^3\right]+\frac{1}{4} C\, \text{Tr}\left[Q^4\right]+(\nabla Q)^2 \,.
\end{eqnarray}

In  the  most general sense it is  a  symmetric traceless  second rank tensor, which vanishes in the symmetric isotropic phase, and presents  a finite value in the ordered nematic phase. In this paper, we parameterize Q as
\begin{equation}
Q=\frac{1}{\sqrt{2}}\left(
\begin{array}{ccc}
 -\frac{\phi _1}{\sqrt{3}}-\phi _2 & \phi _3                            & \phi _4 \\
 \phi _3                             & \phi _2-\frac{\phi _1}{\sqrt{3}} & \phi _5 \\
 \phi _4                             & \phi _5                            & \frac{2 \phi _1}{\sqrt{3}} \\
\end{array}
\right) \,.
\label{Q}
\end{equation}

The central concept of the NPRG formalism is the scale dependent effective average action $\Gamma _k$. In the so-called derivative expansion, it is expanded in powers of the derivative of the order parameter.

\begin{eqnarray}
\Gamma_k[\phi] = \int \mathrm{d}^D x [ U_k(\phi ) + Z_k(\partial \phi)^2 ] \,,
\end{eqnarray}
where $U_k(\phi )$ is the effective average potential and $Z_k$ is the running wavefunction renormalization factor.

The evolution of $\Gamma _k$ with respect to the scale obeys an exact flow equation, the Wetterich equation:
\begin{eqnarray}
\partial_t\Gamma_k[\phi] = \frac 12 \text{Tr} \int \mathrm{d}^D q \partial_k R_k(q^2) ( \Gamma_k^{(2)}[q, -q, \phi] + R_k(q^2) )^{-1} \,,
\end{eqnarray}
where $t=- \text{ln} \frac{k}{k_0}$ (following the convention of \cite{Kopietz:2010zz}), $G_k(q, -q, \phi )$ stands for the propagator matrix. The initial condition is given by the bare action $\Gamma _{k=k_0}(\phi )=S_{k_0}(\phi )$.

$R_k(q^2)$  is the Litim regulator~\cite{Litim:2001up}:
\begin{equation}
R_k(q^2) = Z_k(k^2-q^2) \Theta(k^2-q^2) \,,
\end{equation}
where $Z_k$ is the running wave function renormalization and  $\Theta$ is the usual step function.

The Lagrangian density can be expanded in terms of two basic invariants:
\begin{eqnarray}
\rho = \text{Tr}\left[Q^2\right] &=& \phi _1^2+\phi _2^2+\phi _3^2+\phi _4^2+\phi _5^2 \,, \nonumber\\
\tau =\text{Tr}\left[Q^3\right]  &=& \frac{1}{6 \sqrt{2}}\Big(2 \sqrt{3} \phi _1^3-9 \phi _2 \phi _4^2+9 \phi _2 \phi _5^2+18 \phi _3 \phi _4 \phi _5 \nonumber\\
&+& 3 \sqrt{3}  \phi _1\left(-2 \phi _2^2-2 \phi _3^2 + \phi _4^2+\phi _5^2\right)\Big) \,,
\end{eqnarray}
$\rho$ and $\tau$ are not totally independent. There exists an interesting  property for these invariants which will be quite useful in the following calculations:
\begin{eqnarray}
6 \tau ^2-\rho ^3=-\left(\phi _2^3-3 \phi _1^2 \phi _2\right){}^2\leq 0 \,.
\label{property}
\end{eqnarray}

We consider the following  effective average action for Landau-de Gennes model:
\begin{eqnarray}
\Gamma _k(\phi )=\int d^Dx\left(U_k(\rho ,\tau )+Z_k \left(\left(\partial \phi _1\right){}^2+\ldots +\left(\partial \phi _5\right){}^2\right)\right) \,,
\end{eqnarray}
where  $\Phi$ a constant field configuration. Since the order parameter is symmetric, we can always find a frame of reference to diagonalize Q. So we simply set the nondiagonal terms $\phi _3$, $\phi _4$, $\phi _5$ to be zero:
\begin{eqnarray}
\Phi =\frac{1}{\sqrt{2}}\left(
\begin{array}{ccc}
 -\frac{\phi _1}{\sqrt{3}}-\phi _2 & 0 & 0 \\
 0 & \phi _2-\frac{\phi _1}{\sqrt{3}} & 0 \\
 0 & 0 & \frac{2 \phi _1}{\sqrt{3}} \\
\end{array}
\right) \,.
\label{QPhi}
\end{eqnarray}
The flow equation for the effective average potential follows from its definition:
\begin{eqnarray}
U_k(\rho ,\tau )=\frac{1}{\Omega}\Gamma _k(\phi )|_{\Phi } \,,
\end{eqnarray}
where $\Omega$ is the volume of the system and $\Phi$ is the above field configuration.

There also exists a microscopic description of the liquid crystals, the simplest of which are rigid rods. Taking the z-axis of laboratory frame as the axis of ordering, The order parameter reads
\begin{eqnarray}
Q^M=S \left(
\begin{array}{ccc}
 -\frac{1}{3} & 0 & 0 \\
 0 & -\frac{1}{3} & 0 \\
 0 & 0 & \frac{2}{3} \\
\end{array}
\right) \,,
\label{Qz}
\end{eqnarray}
while in an arbitrary frame of reference:
\begin{eqnarray}
Q_{\alpha\beta}^M = S (n_\alpha n_\beta - \tfrac13 \delta_{\alpha\beta}) \,,
\end{eqnarray}
where $S=\frac{3}{2} \left\langle\cos ^2\theta \right\rangle -\frac{1}{2}$, $\hat{n}$ describes the average direction of the alignment of molecules (i.e. the direction of the nematic axis). $Q^M$ is equal to Eq.~(\ref{QPhi}) provided $\phi _2=0$, $\phi _1=\sqrt{\frac{2}{3}} S$.

The two basic invariants  $\rho$ and $\tau$ reads
\begin{eqnarray}
\phi _1^2+\phi _2^2=\rho \,,\\
\frac{\phi _1 \left(\phi _1^2-3 \phi _2^2\right)}{\sqrt{6}}=\tau \,.
\end{eqnarray}
We can then solve the field in terms of the invariants
\begin{eqnarray}
\phi _1=-\frac{1}{2} \left(\frac{\sqrt[3]{ \sqrt{36 \tau ^2-6\rho ^3}-6 \tau }}{\sqrt[6]{6}}+\frac{\sqrt[6]{6} \rho }{\sqrt[3]{ \sqrt{36 \tau ^2-6\rho ^3}-6 \tau }}\right) \,,
\end{eqnarray}
\begin{widetext}
\begin{eqnarray}
\phi _2=-\frac{1}{2 \sqrt{6}}\sqrt{ -\left( 6\sqrt{36 \tau ^2-6\rho ^3}-36 \tau \right)^{2/3}-\frac{6 \sqrt[3]{6} \rho ^2}{\left( \sqrt{36 \tau ^2-6\rho ^3}-6 \tau \right)^{2/3}}+12 \rho } \,.
\end{eqnarray}
\end{widetext}
There are also 5 other solutions, but each of them lead to the same final flow equations. Note that the property Eq.~(\ref{property}) can be used to transform the solution into complex functions.  For example
\begin{equation}
\sqrt{36 \tau ^2-6\rho ^3}-6 \tau =-\sqrt{6} \rho ^{\tfrac 32} \exp \left(-i \tan ^{-1}(\frac{ \sqrt{\rho ^3-6 \tau ^2}}{\sqrt{6} \tau })\right) \,.
\end{equation}
When we substitute the complex function back to the solutions, the imaginary parts all exactly cancel out, leaving the real part, as it should be, for example,
\begin{eqnarray}
\phi _1 &=& -\sqrt{\rho } \sin \left(\frac{1}{3} \tan ^{-1}\left(\frac{\sqrt{\rho ^3-6 \tau ^2}}{\sqrt{6} \tau }\right)+\frac{\pi }{6}\right) \,, \\
\phi _2&=& -\frac{1}{\sqrt{2}}\sqrt{\rho  \sin \left(\frac{\pi }{6}-\frac{2}{3} \tan ^{-1}\left(\frac{\sqrt{\rho ^3-6 \tau ^2}}{\sqrt{6} \tau }\right)\right)+ \rho } \,. \nonumber
\end{eqnarray}
Then we can calculate all the two point vertices (and the propagator matrix as its inverse) and three point vertices term by term. The flow equation for the potential reads
%
\begin{equation}
\frac{\partial U_k(\rho ,\tau )}{\partial k}=\frac{k^{D-1} K_D}{D} Z_k k^2 \left(1-\frac{\eta _k}{D+2}\right) \left(\Sigma_{i=1}^5 \frac{1}{ Z_k k^2 +M_i^2}\right) \,,
\end{equation}
%
where $K_D=\frac{1}{2^{D-1} \pi ^{D/2} \Gamma \left(\frac{D}{2}\right)}$
and the masss eigenvalues are given by
\begin{widetext}
\begin{eqnarray}
M_1^2 &=& \frac{1}{2} \left(\sqrt{\rho } \left(3 \sqrt{\cos \left(\frac{\alpha }{3}\right)+1}-\sqrt{6} \sin \left(\frac{\alpha }{6}\right)\right) U_k{}^{(0,1)}(\rho ,\tau )+4 U_k{}^{(1,0)}(\rho ,\tau )\right) \,,\\
M_2^2 &=& \frac{1}{2} \left(-\sqrt{\rho } \left(3 \sqrt{\cos \left(\frac{\alpha }{3}\right)+1}+\sqrt{6} \sin \left(\frac{\alpha }{6}\right)\right) U_k{}^{(0,1)}(\rho ,\tau )+4 U_k{}^{(1,0)}(\rho ,\tau )\right) \,,\\
M_3^2&=&\sqrt{6 \rho } \sin \left(\frac{\alpha }{6}\right) U_k{}^{(0,1)}(\rho ,\tau )+2 U_k{}^{(1,0)}(\rho ,\tau ) \,,\\
M_4^2&=&p-\frac{\sqrt{q}}{4} \, \,\,\, ,   M_5^2=p+\frac{\sqrt{q}}{4} \,,
\end{eqnarray}
\end{widetext}
%
%
where $\alpha =2 \tan ^{-1}\left(\frac{\sqrt{\rho ^3-6 \tau ^2}}{\sqrt{6} \tau }\right)+\pi$, and
%
\begin{widetext}
\begin{eqnarray}
p&=&\frac{3}{4} \rho ^2 U_k{}^{(0,2)}(\rho ,\tau )+2 \rho  U_k{}^{(2,0)}(\rho ,\tau )+2 U_k{}^{(1,0)}(\rho ,\tau )+6 \tau  U_k{}^{(1,1)}(\rho ,\tau ) \,, \\
q&=&96 \rho  U_k{}^{(0,1)}(\rho ,\tau )^2+48 \left(4 \rho ^2 U_k{}^{(1,1)}(\rho ,\tau )+3 \rho  \tau  U_k{}^{(0,2)}(\rho ,\tau )+8 \tau  U_k{}^{(2,0)}(\rho ,\tau )\right) U_k{}^{(0,1)}(\rho ,\tau ) \nonumber\\
&+&\rho  \Big\{48 \rho  U_k{}^{(0,2)}(\rho ,\tau ) \left(\rho  U_k{}^{(2,0)}(\rho ,\tau ) \cos \left(2 \tan ^{-1}\left(\frac{\sqrt{\rho ^3-6 \tau ^2}}{\sqrt{6} \tau }\right)\right)+3 \tau  U_k{}^{(1,1)}(\rho ,\tau )\right) \nonumber\\
&+&9 \rho ^3 U_k{}^{(0,2)}(\rho ,\tau )^2+32 \left(3 \rho ^2 U_k{}^{(1,1)}(\rho ,\tau )^2+12 \tau  U_k{}^{(2,0)}(\rho ,\tau ) U_k{}^{(1,1)}(\rho ,\tau )+2 \rho  U_k{}^{(2,0)}(\rho ,\tau )^2\right)\Big\} \,.
\end{eqnarray}
\end{widetext}
It is often convenient to work with dimensionless renormalized quantities that are defined by $\tilde{\rho }=Z_k k^{2-D} \rho $, $\tilde{\tau }= \left( Z_k k^{2-D}\right){}^{3/2} \tau$, $\tilde{U}_t\left(\tilde{\rho },\tilde{\tau }\right)=k^{-D} U_k(\rho ,\tau )$.

The flow equation for the dimensionless  potential reads

\begin{widetext}
\begin{eqnarray}
\frac{\partial \tilde{U}_t\left(\tilde{\rho },\tilde{\tau }\right)}{\partial t}=\tilde{\rho } \left(-D-\eta_t \tilde{\rho} +2\right) \frac{\partial \tilde{U}_t\left(\tilde{\rho },\tilde{\tau }\right)}{\partial \tilde{\rho }}+\frac{3}{2} \tilde{\tau } \left(-D-\eta_t \tilde{\tau}+2\right) \frac{\partial \tilde{U}_t\left(\tilde{\rho },\tilde{\tau }\right)}{\partial \tilde{\tau }}+D \tilde{U}_t\left(\tilde{\rho },\tilde{\tau }\right)  \nonumber\\
- \frac{ K_D }{D} \left(1-\frac{\eta _t}{D+2}\right)\left(\frac{1}{m_1^2+1}+\frac{1}{m_2^2+1}+\frac{1}{m_3^2+1}+\frac{1}{m_4^2+1}+\frac{1}{m_5^2+1}\right) \,,
\end{eqnarray}
\end{widetext}
where $m_i^2$  are the dimensionless counterpart of $M_i^2$. For example
\begin{widetext}
\begin{equation}
m_1^2=\frac{1}{2} \left(\sqrt{\tilde{\rho }} \left(3 \sqrt{\cos \left(\frac{\tilde{\alpha }}{3}\right)+1}-\sqrt{6} \sin \left(\frac{\tilde{\alpha }}{6}\right)\right) \tilde{U}_t{}^{(0,1)}\left(\tilde{\rho },\tilde{\tau }\right)+4 \tilde{U}_t{}^{(1,0)}\left(\tilde{\rho },\tilde{\tau }\right)\right) \,.
\end{equation}
\end{widetext}

This is the full  equation for the  potential. But it is extremely difficult to solve directly due to its complexity. In order to extract useful physics from it, we make the following approximations.
To simplify the notation we introduce
\begin{eqnarray}
\epsilon (\tilde{\rho } )=\tilde{U}_t\left[\tilde{\rho },\tilde{\tau }=0\right] \,,\\
\epsilon '(\tilde{\rho } )=\tilde{U}_t{}^{(1,0)}\left[\tilde{\rho },\tilde{\tau }=0\right] \,,\\
\lambda (\tilde{\rho } )=3\tilde{U}_t{}^{(0,1)}\left[\tilde{\rho },\tilde{\tau }=0\right] \,.
\end{eqnarray}

For $n\geq 2$, we set  $\tilde{U}_t{}^{(0,n)}\left[\tilde{\rho },\tilde{\tau }=0\right]=0$, then $\tilde{U}_t\left(\tilde{\rho },\tilde{\tau }\right)$ can be expressed as
\begin{equation}
\tilde{U}_t\left(\tilde{\rho },\tilde{\tau }\right)=\frac{1}{3} \tilde{\tau } \lambda (\tilde{\rho } )+\epsilon (\tilde{\rho } ) \,.
\end{equation}

The scale dependence of $\epsilon$ is given by

\begin{widetext}
\begin{eqnarray}
\frac{\partial \epsilon }{\partial t} &=& D \epsilon -D \tilde{\rho } \epsilon '-\tilde{\rho } \eta _t \epsilon '+2 \tilde{\rho } \epsilon ' -\frac{K_D }{D} \left(1-\frac{\eta _t}{D+2}\right)
\Big( \frac{1}{2 \epsilon '+1} + \frac{8 \epsilon '+4}{-\lambda ^2 \tilde{\rho }+8 \epsilon '^2+8 \epsilon '+2}  \nonumber\\
&& +\frac{6 \left(2 \tilde{\rho } \epsilon ''+2 \epsilon '+1\right)}{-2 \lambda ^2 \tilde{\rho }-2 \tilde{\rho }^3 \lambda '^2-4 \lambda  \tilde{\rho }^2 \lambda '+12 \tilde{\rho } \epsilon ''+12 \epsilon ' \left(2 \tilde{\rho } \epsilon ''+1\right)+12 \epsilon '^2+3} \Big) \,.
\end{eqnarray}
Similarly for $\lambda$ one finds
\begin{eqnarray}
\frac{\partial \lambda }{\partial t} &=& D \lambda -3 \tilde{\rho } \lambda ' \left(D+\eta _t-2\right) -\frac{3}{2} \lambda  \left(D+\eta _t-2\right) -\frac{ K_D }{D} \left(1-\frac{\eta _t}{D+2}\right) \Big( \frac{4 \left(\lambda -6 \tilde{\rho } \lambda '\right) \left(\lambda ^2 \tilde{\rho }+8 \epsilon '^2+8 \epsilon '+2\right)}{\tilde{\rho } \left(-\lambda ^2 \tilde{\rho }+8 \epsilon '^2+8 \epsilon '+2\right)^2} \nonumber \\
&& +\frac{4 \left(\tilde{\rho } \lambda ''+4 \lambda '\right)}{-2 \lambda ^2 \tilde{\rho }-2 \tilde{\rho }^3 \lambda '^2-4 \lambda  \tilde{\rho }^2 \lambda '+12 \tilde{\rho } \epsilon ''+12 \epsilon ' \left(2 \tilde{\rho } \epsilon ''+1\right)+12 \epsilon '^2+3} -\frac{2 \left(3 \tilde{\rho } \lambda '+\lambda \right)}{\tilde{\rho } \left(2 \epsilon '+1\right)^2} \nonumber \\
&& -\frac{12 \left(\sqrt{2} \tilde{\rho }^{3/2} \lambda '+ \lambda  \sqrt{2 \tilde{\rho }}-2 \tilde{\rho } \epsilon ''\right) \left(\epsilon '' \left(2 \tilde{\rho } \lambda '-6 \lambda \right)+\left(2 \epsilon '+1\right) \left(\tilde{\rho } \lambda ''+4 \lambda '\right)\right)}{\left(-2 \lambda ^2 \tilde{\rho }-2 \tilde{\rho }^3 \lambda '^2-4 \lambda  \tilde{\rho }^2 \lambda '+12 \tilde{\rho } \epsilon ''+12 \epsilon ' \left(2 \tilde{\rho } \epsilon ''+1\right)+12 \epsilon '^2+3\right)^2}\Big) \,.
\label{lam}
\end{eqnarray}
\end{widetext}

The evolution equation for  its derivative $\epsilon '(\tilde{\rho })$ can also be calculated analogously:
\begin{widetext}
\begin{eqnarray}
\frac{\partial \epsilon '}{\partial t} &=& D \epsilon ' -\tilde{\rho } \epsilon '' \left(D+\eta _t-2\right)-\epsilon ' \left(D+\eta _t-2\right) -\frac{K_D}{D} \left(1-\frac{\eta _t}{D+2}\right)
\Big(\frac{4 \lambda  \left(2 \epsilon '+1\right) \left(2 \tilde{\rho } \lambda '+\lambda \right)-8 \epsilon '' \left(\lambda ^2 \tilde{\rho }+2 \left(2 \epsilon '+1\right)^2\right)}{\left(\lambda ^2 \tilde{\rho }-2 \left(2 \epsilon '+1\right)^2\right)^2} \nonumber \\
&& +\frac{12 \left(\tilde{\rho } \epsilon '''+2 \epsilon ''\right)}{-2 \lambda ^2 \tilde{\rho }-2 \tilde{\rho }^3 \lambda '^2-4 \lambda  \tilde{\rho }^2 \lambda '+12 \tilde{\rho } \epsilon ''+12 \epsilon ' \left(2 \tilde{\rho } \epsilon ''+1\right)+12 \epsilon '^2+3}  -\frac{2 \epsilon ''}{\left(2 \epsilon '+1\right)^2} \\
&&-\frac{12 \left(2 \tilde{\rho } \epsilon ''+2 \epsilon '+1\right) \left(-2 \lambda  \tilde{\rho } \left(\tilde{\rho } \lambda ''+3 \lambda '\right)+12 \tilde{\rho } \epsilon ''^2+\tilde{\rho } \left(6 \left(2 \epsilon '+1\right) \epsilon '''-\tilde{\rho } \lambda ' \left(2 \tilde{\rho } \lambda ''+5 \lambda '\right)\right)-\lambda ^2+12 \left(2 \epsilon '+1\right) \epsilon ''\right)}{\left(-2 \lambda ^2 \tilde{\rho }-2 \tilde{\rho }^3 \lambda '^2-4 \lambda  \tilde{\rho }^2 \lambda '+12 \tilde{\rho } \epsilon ''+12 \epsilon ' \left(2 \tilde{\rho } \epsilon ''+1\right)+12 \epsilon '^2+3\right)^2} \Big) \nonumber
\label{epsilon}
\end{eqnarray}
\end{widetext}

It is worth noting that if our only objective is to get the above simplified equations rather than the full machinery, we can also circumvent the substitution technique, and set $\tilde{\tau }=0$ directly at last. We have verified that these two approaches indeed lead to the same equations.

$Z_k$ can be extracted from the flow equation through:
\begin{eqnarray}
Z_k=\frac{(2 \pi )^D}{\delta (0)}\lim_{p^2\to 0} \frac{\mathrm d}{\mathrm d p^2}\left(\frac{\partial ^2\Gamma _k}{\delta \phi(p)  \delta \phi  (-p)}\right) \,.
\end{eqnarray}

The anomalous dimension is defined by
\begin{eqnarray}
\eta _t=-\frac{k}{Z_k} \frac{\partial Z_k}{\partial k} \,.
\end{eqnarray}

It can be shown that:
\begin{eqnarray}
\eta _{m n} &=& -\frac{k}{Z_{m n}} \frac{\partial Z_{m n}}{\partial k} \\
&=& \frac{k^{D+2} K_D }{D Z_{m n}} \Sigma\left[Z_{a b} G_{b c}  \Gamma _{c d n} G_{d i}Z_{i j}   G_{j e}\Gamma _{e f m}  G_{f a}  \right] \,. \nonumber
\end{eqnarray}
For our model, $Z_{m n}=Z_k \delta _{\text{mn}}$. $G_{\text{ij}}$ is the propagator matrix element, where the propagator matrix is defined by the inverse of the two point vertex. And $\Gamma _{c d n}$ is the three point vertex.

In this paper, without loss of generality we take m=4 and n=4, and get  the  anomalous dimension:
\begin{widetext}
\begin{eqnarray}
\eta _t &=& \frac{K_D }{D} \Big(\frac{4 \lambda ^2}{\left(\lambda ^2 \tilde{\rho }-8 \epsilon '^2-8 \epsilon '-2\right)^2} +\frac{24 \left(2 \lambda  \tilde{\rho } \lambda '+\tilde{\rho }^2 \lambda '^2+\lambda ^2-3 \epsilon ''-6 \epsilon ' \epsilon ''\right)}{\left(2 \lambda ^2 \tilde{\rho }+2 \tilde{\rho }^3 \lambda '^2+4 \lambda  \tilde{\rho }^2 \lambda '-12 \tilde{\rho } \epsilon ''-24 \tilde{\rho } \epsilon ' \epsilon ''-12 \epsilon '^2-12 \epsilon '-3\right)^2} \nonumber \\
&& -\frac{4 \left(-2 \lambda  \tilde{\rho } \lambda '-\tilde{\rho }^2 \lambda '^2-\lambda ^2+6 \epsilon ''+12 \epsilon ' \epsilon ''\right)}{\left(2 \epsilon '+1\right)^2 \left(2 \lambda ^2 \tilde{\rho }+2 \tilde{\rho }^3 \lambda '^2+4 \lambda  \tilde{\rho }^2 \lambda '-12 \tilde{\rho } \epsilon ''-24 \tilde{\rho } \epsilon ' \epsilon ''-12 \epsilon '^2-12 \epsilon '-3\right)}  \Big) \,.
\end{eqnarray}
\end{widetext}
When $\lambda =0$ and evaluated at the minimum $\tilde{\rho }_0$ (i.e. $\epsilon '(\tilde{\rho }_0)=0$) this reduces to
\begin{equation}
\eta_t=\frac{32 \tilde{\rho }_0 K_D \epsilon ''^2}{D \left(4 \tilde{\rho }_0 \epsilon ''+1\right){}^2} \,,
\end{equation}
which is identical to the usual $O(5)$ model expression, if considering the different definition of $\rho$.

\section{Numerical Results}
\label{sec-results}

We solve the combined equations Eq.~(\ref{lam}) and Eq.~(\ref{epsilon}) together~\cite{Adams:1995cv} by discretizing the differential equations in this form:
\begin{equation}
\frac{u_i(\text{$\Delta $t}+t)-u_i(t)}{\text{$\Delta $t}}=F\left(u_i(\text{$\Delta $t}+t),u_i^{(1)}(\text{$\Delta $t}+t),u_i^{(2)}(\text{$\Delta $t}+t)\right) \,,
\end{equation}
where $u_i(t)$(i from 1 to n) denotes $\epsilon '(\tilde{\rho })$, $u_i(t)$(i from n+1 to 2n) denotes $\lambda(\tilde{\rho })$ and F is the discretized righthand side of the combined equations.
To evaluate  the derivatives  on the righthand side we use 6-point formulas.
We take 60 points for the discretization of the variable (i.e. n=60). We solve these nonlinear algebraic equations using the Newton Raphson algorithm. The basic idea is to set the previous solution as the trial solution and searching for the next step solution with the aid of Jacobian  many times until the necessary accuracy is achieved.

We solve the above equation for the first few steps, and then switch to the following equation to improve its accuracy:
\begin{eqnarray}
\frac{1}{12 \text{$\Delta $t}} \big\{25 u_i (\text{$\Delta $t}+t)-48 u_i(t)+36 u_i (t-\text{$\Delta $t})-16 u_i (t-2 \text{$\Delta $t}) \nonumber \\
+3 u_i (t-3 \text{$\Delta $t})\big\} = F\left(u_i(\text{$\Delta $t}+t),u_i^{(1)}(\text{$\Delta $t}+t),u_i^{(2)}(\text{$\Delta $t}+t)\right) \,.
\end{eqnarray}

\begin{figure}[!htb]
\begin{center}
\includegraphics[width=240pt]{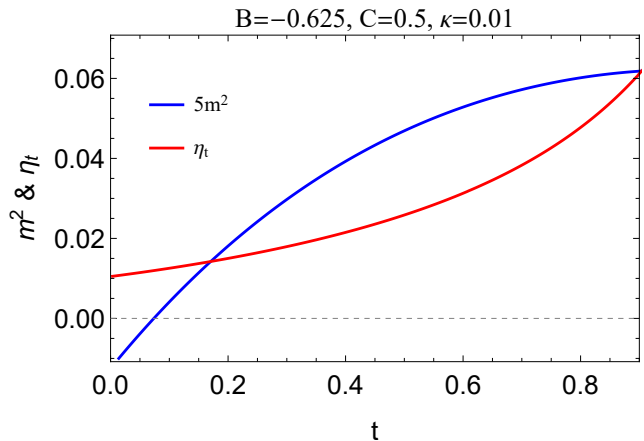}
\caption{Evolution of the renormalized mass and anomalous dimension with respect to RG time t.}
\label{fig-plotrman}
\end{center}
\end{figure}

\begin{figure}[!htb]
\begin{center}
\includegraphics[width=240pt]{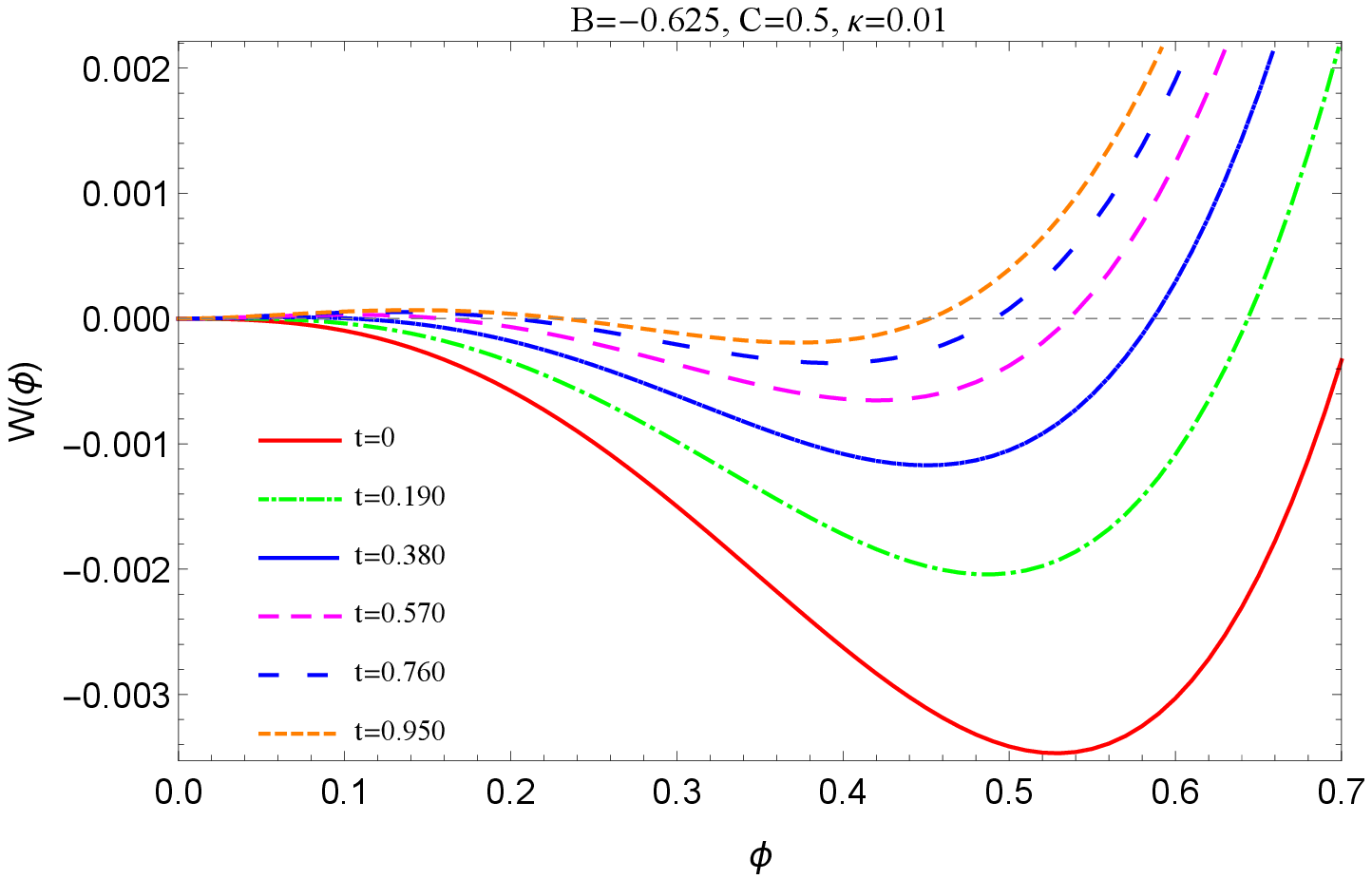}
\caption{Evolution of the effective average potential with respect to the field variable at different RG time t.}
\label{fig-plot2}
\end{center}
\end{figure}

\begin{figure}[!htb]
\begin{center}
\includegraphics[width=240pt]{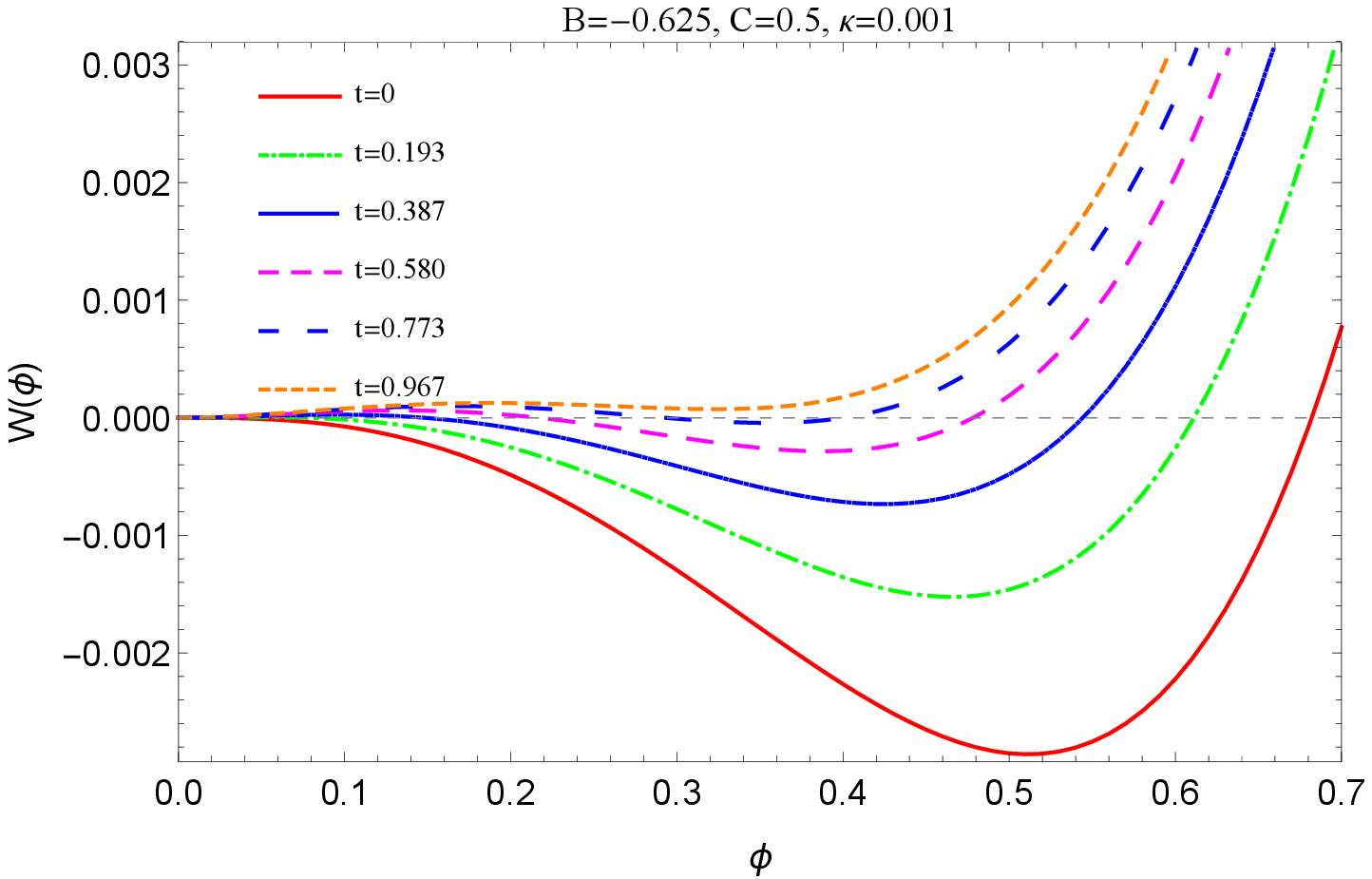}
\caption{Evolution of the effective average potential with respect to the field variable at different RG time t.}
\label{fig-plot5}
\end{center}
\end{figure}

\begin{figure}[!htb]
\begin{center}
\includegraphics[width=240pt]{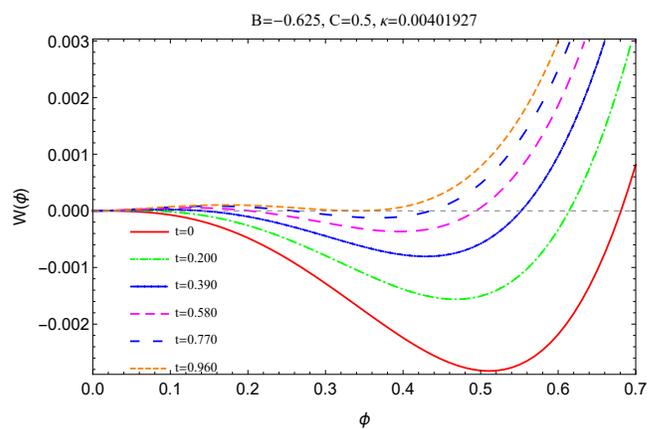}
\caption{Evolution of the effective average potential  at different RG time t for critical $\kappa$.}
\label{fig-plotc}
\end{center}
\end{figure}

\begin{figure}[!htb]
\begin{center}
\includegraphics[width=240pt]{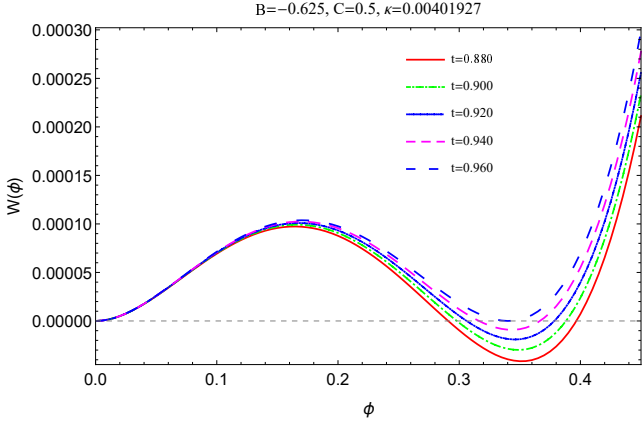}
\caption{Enlarged portion of Fig.~\ref{fig-plotc}.}
\label{fig-plotcl}
\end{center}
\end{figure}

\begin{figure}[!htb]
\begin{center}
\includegraphics[width=240pt]{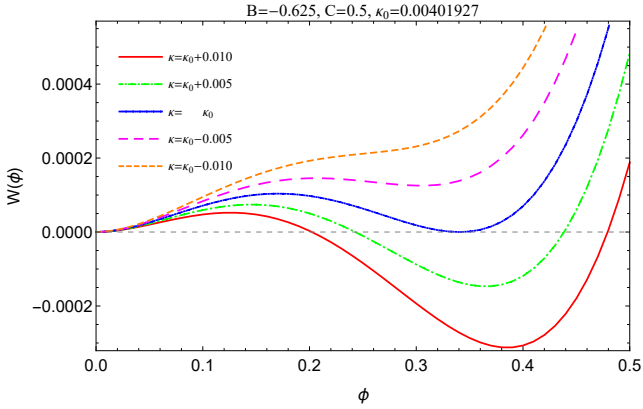}
\caption{Evolution of the effective average potential,we can see clearly the jump of the unrenormalized order parameter. The middle touching curve corresponds to the one in Fig.~\ref{fig-plotcl}. }
\label{fig-plotphtr}
\end{center}
\end{figure}

\begin{figure}[!htb]
\begin{center}
\includegraphics[width=240pt]{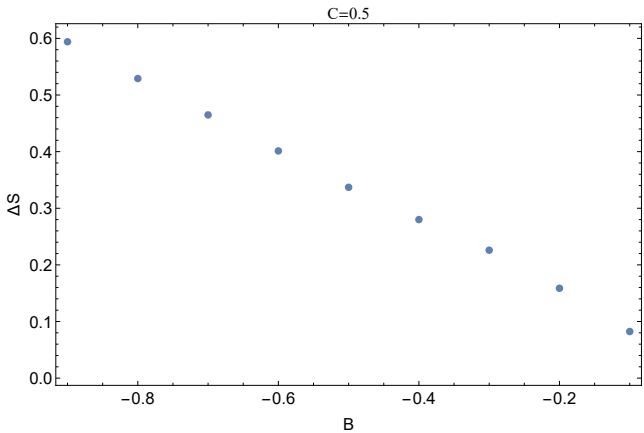}
\caption{The relationship between order parameter jump $\text{$\Delta $S}$ and the cubic term.}
\label{fig-plotBS}
\end{center}
\end{figure}

When B=0, we can reproduce the results of O(5) model. Then we turn on the explicit symmetry breaking term.
It is convenient to define $\kappa=-\frac{A}{C}$ and rewrite the initial potential as
\begin{equation}
U(\rho ,\tau )=\frac{B \tau }{3}+\frac{1}{4} C (\rho -\kappa )^2=\frac{B \tau }{3}+\frac{C \kappa ^2}{4}-\frac{C \kappa  \rho }{2}+\frac{C \rho ^2}{4} \,,
\end{equation}
differing by an irrelevant constant term.
For specific values of the parameters, we first get the evolution result for $\epsilon '(\tilde{\rho } )$ ($\epsilon \left(\tilde{\rho }\right)$ can be calculated through direct integration) and $\lambda \left(\tilde{\rho }\right)$. We then transform back to the original unscaled variables. To get the physical effective potential, we project our solution to the branch $\phi _2=0$, $\rho =\phi _1^2$, $\tau =\frac{\phi _1^3}{\sqrt{6}}$, i.e. $\tau =\frac{\rho ^{3/2}}{\sqrt{6}}$. We omit the subscript and  denote resulting effective potential as $W(\phi)$. Part of the results  are shown in the figures. We stop the flow when the renormalized mass term $m^2$  approaches a constant (Fig.~\ref{fig-plotrman}). The stopping RG time t is  roughly the same for different initial $\kappa$. The running anomalous dimension (Fig.~\ref{fig-plotrman}) is evaluated at the minimum $\phi _0$ of $W(\phi)$. In Fig.~\ref{fig-plot2} we show a typical flow of the running effective average potential. The effective potential at the end of the flow lies in the nematic phase. While the potential for a smaller $\kappa$ (Fig.~\ref{fig-plot5}) flows into a symmetric isotropic phase. Between these two $\kappa$ we can find a critical $\kappa$ value that corresponds  exactly to the point where the first order phase transition happens (Fig.~\ref{fig-plotc}, see Fig.~\ref{fig-plotcl} for an enlarge version). This is achieved through a bisection algorithm. From Fig.~\ref{fig-plotcl} and Fig.~\ref{fig-plotphtr}  we can read off the jump of the order parameter. For different values of B the order parameters are mapped out following above procedures (Fig.~\ref{fig-plotBS}). The order parameter jump decreases with increasing values of B. For different parameters of C we can get  similar results.


\begin{figure}[!htb]
\begin{center}
\includegraphics[width=240pt]{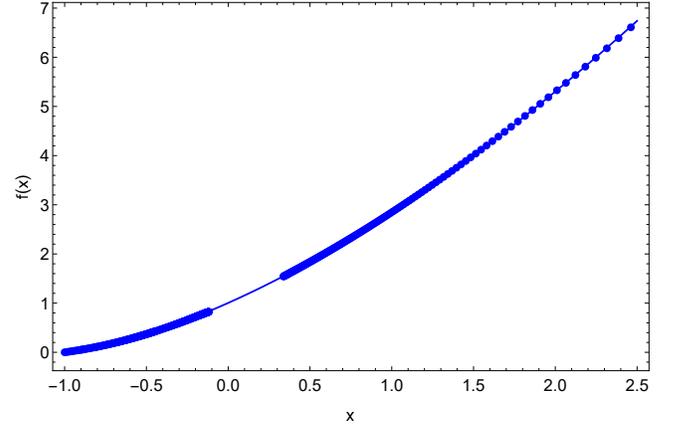}
\caption{Scaling function for the equation of state without the cubic term.}
\label{fig-EOS}
\end{center}
\end{figure}

We also calculate the NI transition temperature difference to shed some new light on the NI puzzle~\cite{Mukherjee1998, Mukherjee1995, Mukherjee1994, Priest:1978}. We follow similar techniques used in \cite{Mukherjee1994}. Since the first order phase transition is weak and quite close to the hypothetical  critical point (if there were no cubic term), as a first approximation it is reasonable to treat $\frac{t_0}{\phi ^{1/\beta }}$ and $\frac{B}{\phi ^{\omega }}$ as two scaling variables to get the equation of state~\cite{Braun:2007td, Berges:1996ja}. According to $\frac{W'(\phi)}{\phi ^{\delta}}=f(x)$, we get(Fig.~\ref{fig-EOS})
\begin{eqnarray}
f(x)=a (x+1)^b \left(c(x+1)+d \sqrt{x+1}+1\right) \,,
\end{eqnarray}
where $x=\frac{t_0}{\phi ^{1/\beta }}$, $t_0 =\frac{T_c-T^*}{T^*}$, $\beta =\frac{1}{2} \nu  (d+\eta -2)$, $\delta =\frac{D-\eta +2}{D+\eta -2}$, $\nu =0.8377$, $\eta =0.0377$, $\beta =0.4347$, $\delta =4.781$, $a=0.253$, $b=1.076$, $c=0.221$, $d=2.748$.

We have verified the above scaling relations to high accuracy. As a preliminary treatment we then get the equation of state for Landau-de Gennes Model:
\begin{eqnarray}
\frac{H}{\phi ^{\delta }}+\frac{g B}{\phi ^{\omega }}=f(x) \,,
\end{eqnarray}
where H is the external field and $B\sim \text{$\Delta $S}^{\omega }$. We have  determined $g=0.67$, $\omega=1.10$(for similar result see~\cite{Berges:1996ja}).\\
From thermodynamic arguments we can obtain the free energy by integrating the equation of state with respect to $\phi$. Then we can express the conditions that the free energy be equal for the nematic state  and that the free energies be a local minimum with respect to $\phi$ as follows
\begin{subequations}
\begin{eqnarray}
\int_0^{\phi _c} H\left(\phi '\right) \, d\phi '=0 \,,
\end{eqnarray}
\begin{eqnarray}
H(\phi _c)=0 \,,
\end{eqnarray}
\end{subequations}
where $\phi _c=\sqrt{\frac{2}{3}} S_c$, the experimental value $S_c=0.4$. We then solve the equations and get $t_{0c}=0.0195$, which corresponds to the temperature difference $T_c-T^*=5.85 K$. Although still far from the result of experiment, this result is close to the result of~\cite{Mukherjee1994}, and much smaller than the mean field value. This indicates that we maybe on the right track to finally resolve the NI puzzle.

\section{conclusion}
\label{sec-summary}

In this paper we investigated the Landau-de Gennes model in the framework of functional renormalization group. The Lagrangian density can be expanded in terms of two basic invariant combinations of the elements of the order parameter tensor. We then solve the field variables of the order parameter in terms of the two invariants. When transformed into complex variables, the imaginary parts exactly cancel. With the aid of  Litim regulator  the full analytic flow equation for the potential and its dimensionless counterpart are derived. A truncation is made to simplify the computations and get two coupled partial differential equations for the cubic and quartic ``couplings''. We also derived the flow equation for the anomalous dimension. The two coupled equations are solved on a grid with Newton Raphson method.A large parameter space of the model is mapped and first order phase transitions are observed.

With the experimental value of the order parameter jump as an input, we also obtained the NI transition temperature difference. Our result shows an improvement over previous values. Much more interesting work can be done. For instance, we expect that a more refined and accurate analysis of the equation of state can give a better improvement. Further, our formalism can be easily extended to other similar models including the explicit symmetry breaking term. We leave these for future work.

\section*{Acknowledgement}

The work is supported in part by the Ministry of Science and Technology of China (MSTC) under the ``973" Project Nos. 2015CB856904(4) (DH), and by NSFC under Grant Nos. 11735007 (DH and MH),  11375070,11521064  (DH) and  Nos. 11725523, 11261130311(MH).
H.~Z. gratefully acknowledges financial support from China Scholarship Council (CSC) Grant No.~201706770051.

\end{document}